\documentclass[useAMS,usenatbib,usegraphicx]{mn2e}

\title[Are groups of galaxies gravitationally bound?]
{Are the nearby groups of galaxies gravitationally bound objects?}
\author[Sami-Matias Niemi et al.]{Sami-Matias Niemi,$^{1}$\thanks{E-mail:saniem@utu.fi (SMN)}
Pasi Nurmi,$^{1}$\thanks{E-mail:pasnurmi@utu.fi}
Pekka Hein\"am\"aki,$^{1}$\thanks{E-mail:pekheina@utu.fi} and
Mauri Valtonen$^{1}$\thanks{E-mail:mvaltonen2001@yahoo.com}\\
$^{1}$University of Turku, Tuorla Observatory, V\"ais\"al\"antie 20, Piikki\"o, Finland}
 
\begin{document}

\date{Released 2007}

\pagerange{\pageref{firstpage}--\pageref{lastpage}} \pubyear{200x}

\maketitle

\label{firstpage}

%ABS about 200 words
\begin{abstract}
We have compared numerical simulations to observations for the nearby ($< 40$ 
Mpc) groups of galaxies (\citealt{HG82} and \citealt{RGPC}). The group 
identification is carried out using a group-finding algorithm developed by 
\citet{HG82}. Using cosmological N-body simulation code with the $\Lambda$CDM 
cosmology, we show that the dynamical properties of groups of galaxies 
identified from the simulation data are, in general, in a moderate, within 
2$\sigma$, agreement with the observational catalogues of groups of galaxies. 
As simulations offer more dynamical information than observations, we used the 
N-body simulation data to calculate whether the nearby groups of galaxies are 
gravitationally bound objects by using their virial ratio. We show that in 
a $\Lambda$CDM cosmology about $20$ per cent of nearby groups of galaxies, 
identified by the same algorithm as in the case of observations, are not bound, 
but merely groups in a visual sense. This is quite significant, 
specifically because estimations of group masses in observations are often based on 
an assumption that groups of galaxies found by the friends-of-friends algorithm 
are gravitationally bound objects. Simulations with different resolutions show
the same results. We also show how the fraction of gravitationally 
unbound groups varies when the apparent magnitude limit of the sample and the 
value of the cosmological constant $\Lambda$ is changed. In general, a larger 
value of the $\Omega_{\Lambda}$ generates slightly more unbound groups.
\end{abstract}


\section[]{Introduction}\label{s1}

Small groups of galaxies are the most common galaxy associations and contain 
$\sim 50$ per cent of all galaxies in the universe (e.g. \citealt{Holmberg}; 
\citealt{HMS}; \citealt{HG82}; \citealt{GM}; \citealt{Nolt}). The study of 
galaxy groups is a very interesting area of research because these density 
fluctuations lie between galaxies and clusters of galaxies and may provide 
important clues to galaxy formation. Small groups of galaxies are also 
important cosmological indicators of the distribution, and properties, of 
dark matter in the universe.

The dynamics of the nearby groups of galaxies and the Local Group has provided 
a unique challenge to cosmological models in the past. The quiescence of the 
local peculiar velocity field (e.g. \citealt{Vauc}; \citealt{Sand2}; 
\citealt{Sand1} and \citealt{Ek}) was a long-standing puzzle that presented a 
challenge for the models of structure formation. The velocity field within 
5$h^{-1}$ Mpc of the Local Group is extremely 'cold', the dispersion is only 
$\sim 50 - 60$ km s$^{-1}$ (\citealt{Terppa} and references therein). The 
$\Lambda$CDM cosmology and dark energy has solved this problem and it has been 
shown by e.g. \citet{Klypin}, \citet{Maccio} and \citet{PP} with constrained 
simulations that the $\Lambda$CDM cosmology can produce the small values of the 
velocity dispersion. Today, the question of the virialization of groups of 
galaxies and the fraction of gravitationally bound systems provides a new 
challenge for the cosmological models and grouping algorithms.

Over the last two decades, cosmological simulations have proven to be an 
invaluable tool in testing theoretical models in the non-linear regime. The 
standard approach is to assume a cosmological model and to use the appropriate 
power spectrum of the primordial perturbations to construct a random 
realisation of the density field within a given simulation volume. The 
evolution of the initial density field is then followed by using an N-body 
simulation code, and the results in the simulation box (viewed from outside) are 
compared with observational data. A comparison of the simulations with 
observational data is typically done in a statistical manner. The statistical 
approach works well if there is a statistically representative sample of 
objects with well-understood selection effects for both the observed universe 
and the simulations.

Given a set of observed galaxies with their positions in the sky and their 
redshifts, the task of a group-finder is to return sets of galaxies that most 
likely represent true gravitationally bound structures. Some contamination 
is always expected due to selection effects in observations. This study uses 
one of the most popular group-finders: the friends-of-friends (FOF) algorithm. 
FOF has been used widely for identifying groups of galaxies in the redshift 
surveys (\citealt{HG82}; \citealt{GM}; \citealt{Nolt}; \citealt{RGH}; 
\citealt{MFW}; \citealt{RPG}; \citealt{RGPC}; \citealt{Tuck} and 
\citealt{GMCP}) and is now a standard approach.

%Studies of virialization of groups of galaxies are an important method for estimating galaxy masses in observations.
The identification of the group members has, in general, been based on a 
subjective selection of data. In order to remove this difficulty, \citet{HG82} 
(hereafter HG82) developed a method of identifying groups of galaxies from the 
observations. It has been usually thought that the FOF based algorithms would 
produce groupings which are mostly gravitationally bound when the number of 
galaxies in a group exceeds five members (see e.g. \citealt{RGPC}). A few 
studies (e.g. \citealt{Carlberg}) have been made where the free parameters of 
the FOF algorithm have been optimised to avoid spurious groups with 
interlopers. Studies of the FOF algorithms have concluded that the choice of 
the free parameters depends on the case that is studied, and no singular best 
choice of the parameters can be made. Unfortunately, none of the observational 
methods do truly answer the question whether groups of galaxies are 
gravitationally bound objects. There are some estimates for the fraction of how 
many groups found by the FOF algorithm are spurious (see e.g. \citealt{RGPC}), 
but these estimates are not based on the physical properties of the groups.

%In general $\Lambda$CDM -cosmology seems to cope adequately with the challenge of structure formation when few %restrictions are understood.

Cosmological N-body simulations include all the necessary information for 
finding out whether a given object is gravitationally bound or not. 
\cite{Aceves} studied small galaxy groups with N-body simulations and made 
conclusions about the virialization of the groups. \cite{D1994} studied compact 
groups of galaxies with N-body simulations and made remarks about the fraction 
of bound groups and chance alignment systems. Groups of galaxies, and their 
dynamical properties generated by the FOF algorithm, have been studied earlier 
with N-body simulations (see
e.g. %\citealt{F1}; \citealt{F2}; \citealt{DRGF}; Frederic 1995a,b; 
\citealt{NKP}; \citealt{Diaferio}; \citealt{MZ}; \citealt{Luca}). Most of these 
earlier studies have taken advantage of the constrained simulations and have 
used models which are different from the currently popular $\Lambda$CDM 
cosmology. In these constrained, or mock catalogue, simulations the simulated 
space has been compared directly with the observations of redshift space 
and the groups have been categorized as spurious if the algorithm has failed to 
find similar groups as in real space. Instead, we have studied the virial ratio 
of the groups of galaxies produced by the FOF algorithm first described in 
HG82. We will show that $\sim 20$ per cent of the groups of galaxies found by 
the FOF algorithm are not real gravitationally bound groups, but spurious. Our 
results agree roughly with the previous results and estimates but do not 
confirm the claim by \citet{RGH} that groups with more than four members are 
gravitationally bound.

For an 'observer' placed in a specific location, selecting a similar environment
 between observational data and cosmological simulations might be 
problematic. The simplest way is to choose an 'observation' point within a 
simulation box by certain criteria. It is argued (e.g. \citealt{Klypin} and 
references therein) that it is not clear what 'similar environment' actually 
means and that simply placing the observer at some specific point would resolve 
the issue. In this paper, we show that this is a useful approach, as we are not 
comparing simulations to observations directly but with statistics.

The main purpose of this work is to study if groups of galaxies found by the 
HG82 algorithm are bound, and how the fraction of bound groups depend on the chosen 
magnitude limit and cosmological model. We will show our findings with 
different apparent magnitude limits and for different cosmological models. We 
also show that there is no significant correlation between the crossing time
of a group and its virial ratio.

This paper is organized as follows. In section \ref{s2} we review the method 
used in the identification of the group members in the observations. A brief 
discussion of the differences between observations and simulations is given in 
section \ref{s2}. In section \ref{s3} we discuss briefly the virial ratio, used 
for determining whether a group of galaxies is gravitationally bound. Section 
\ref{s4} discusses the simulations we use for the analysis. In sections 
\ref{s5} and \ref{virial} we present our results and discuss the findings. 
Discussion of the probability functions of gravitationally unbound groups is 
done in section \ref{discussion}. Finally, we summarize our results in section 
\ref{summar}.

\section[]{A review of the group-finding algorithm}\label{s2}

In observations there are generally three basic pieces of information available 
for the study of the galaxy distribution: the position, the magnitude and the 
redshift of each galaxy. Although the magnitude is important as a measure of 
the object's visibility, it is usually a poor criterion for group membership. 
The method used for creating a group catalogue in HG82 can be summed up in two 
criteria: the projected separation and the velocity difference. The FOF 
algorithm is described in greater detail by the original authors in HG82.

The grouping method begins with a selection of an object, which has not been 
previously assigned to any of the existing groups. After choosing the object 
the next step is to search for companions with the projected separation 
$D_{12}$ smaller or equal to the separation $D_{L}$:
\begin{equation}
	D_{12} = 2 \sin \left( \frac{\theta}{2} \right) \frac{V}{H_{0}} \lid D_{L}(V_{1}, V_{2}, m_{1}, m_{2}) ~ ,
\end{equation}
where the mean cosmological expansion velocity is 
\begin{equation} 
	V = \frac{V_{1} + V_{2}}{2} ~ , 
\end{equation} 
and the velocity difference $V_{12}$ is smaller or equal to the velocity $V_{L}$: 
\begin{equation} 
	V_{12} = |V_{1} - V_{2}| \lid V_{L}(V_{1}, V_{2}, m_{1}, m_{2}) ~ , 
\end{equation} 
where $V_{1}$ and $V_{2}$ refer to the velocities (redshifts) of the galaxy and 
its companion, $m_{1}$ and $m_{2}$ are their magnitudes, and $\theta$ is their 
angular separation in the sky. If no companions are found, the galaxy is 
entered in a list of isolated galaxies. All companions found are added to the 
list of group members. The surroundings of each companion are then searched by 
using the same method used in the first place to find companions. This process 
is repeated until no further members are found.

There are a variety of prescriptions for $D_{L}$ and $V_{L}$. We adopt the 
method used in HG82, and assume that the luminosity function is independent of 
distance and position and that at larger distances only the fainter galaxies 
are missing. For each pair we take: 
\begin{equation}\label{Dl} 
	D_{L} = D_{0} \left ( \frac{\int_{-\infty}^{M_{12}} \Phi(M)dM}{\int_{- \infty}^{M_{lim}} \Phi(M)dM} \right ) ^{-\frac{1}{3}} ~ , 
\end{equation}
where the integration limits can be calculated from equations: 
\begin{equation}\label{i1}
	M_{lim} = m_{lim} - 25 - 5 \log (D_{F}) 
\end{equation}
and
\begin{equation}\label{i2}
	M_{12} = m_{lim} - 25 - 5 \log(V) ~ , 
\end{equation}
and where $\Phi(M)$ is the differential galaxy luminosity function for the 
sample, and $D_{0}$ is the projected separation in Mpc chosen at some fiducial 
distance $D_{F}$. In this paper we adopt constants $D_{0} = 0.63~\rmn{Mpc}$ and 
$D_{F} = 10~\rmn{Mpc}$ to be same as in HG82. The effect of varying $D_{0}$ has 
been studied e.g. by \citet{RGH} where the effects are also explained.

The limiting velocity difference is scaled in the same way as the distance 
$D_{L}$: 
\begin{equation}\label{Vl}
	V_{L} = V_{0} \left ( \frac{\int_{- \infty}^{M_{12}} \Phi(M)dM}{\int_{- \infty}^{M_{lim}} \Phi(M)dM} \right )^{-\frac{1}{3}} ~ ,
\end{equation}
where the fiducial value is $V_{0} = 400~\rmn{km}~\rmn{s}^{-1}$ and the 
integration limits are as above (Eq. \ref{i1} and \ref{i2}). \cite{RGH} varied 
$V_{0}$ and concluded that the results are not sensitive to the choice of 
$V_{0}$. This is probably related to the geometry of the large-scale structure. 
Frederic (1995a,b) argues that the optimal choice of $D_{0}$ and $V_{0}$ 
depends on the purpose for which groups are being identified. Similar claims 
has been made in papers where the FOF algorithm has been optimized (see e.g. 
\citealt{Eke} and \citealt{Berlind}). Because of this, we also show some 
results when $D_{0} = 0.37~\rmn{Mpc}$ and $V_{0} = 200~\rmn{km}~\rmn{s}^{-1}$ 
are adopted.

Note that the scaling law in Equations \ref{Dl} and \ref{Vl} has been 
questioned by many authors. Specifically, replacing the power -1/3 by -1/2 (see 
the argument in e.g. \citealt{Nolt} and \citealt{Gour}) drastically reduces the 
correlation between the redshift and the velocity dispersion observed in the 
HG82 group catalogue. However, part of this correlation is related to a 
selection effect rather than to the grouping algorithm, because groups with low 
velocity dispersion usually have few bright galaxies and so they can be seen 
only at low redshift. In this paper, we use the equations mentioned above for 
consistency with the HG82 catalogue.

For simplicity, we use the \citet{S} luminosity function: 
\begin{equation}
	\Phi (M) = \frac{2}{5}\Phi^{*} \ln 10 \left( 10^{\frac{2}{5}(M^{*}-M)} 
	\right)^{\alpha + 1} \exp^{-10^{\frac{2}{5}(M^{*}-M)}} ~ , 
\end{equation}
where $M$ is the absolute magnitude of the object. We adopt the parameter 
values of $\alpha = -1.02$, $M^{*} = -19.06$, and $\Phi^{*} = 0.0277$ 
comparable to HG82. For comparison, we use the galaxy luminosity function 
values of $\alpha = -1.15$, $M^{*} = -19.84$, and $\Phi^{*} = 0.0172$, which 
were derived from the Millennium Galaxy Catalogue by \cite{Driver}.

%It is possible to apply the FOF algorithm to a cosmological N-body simulation 
%when few restrictions are kept in mind. The main idea being that the FOF 
%algorithm was designed to function in the redshift space. So when one is to 
%apply the FOF algorithm to a simulation, a simulated real space, coordinates, 
%and real distances must be transformed to a redshift space so the FOF algorithm 
%can function properly.

Our chosen values of constants and parameters needed for the FOF algorithm are 
exactly the same as in HG82 for consistency. We do show selected results with 
more recent values of constants and parameters if these results differ from the 
results produced with the HG82 values. Throughout this paper we adopt the 
parameterized Hubble constant $H_{0} = 
100h~\rmn{km}~\rmn{s}^{-1}~\rmn{Mpc}^{-1}$ with $h=1.0$ for comparison with 
HG82 when a definite value of the Hubble constant is needed. We do not consider 
dust extinction in the analysis of the simulations in this paper.

\section[]{A review of the virial theorem}\label{s3}

In the simplest case when we have a two body system with total mass $M = m_{1} 
+ m_{2}$ and $V$ is the relative speed of the components, the kinetic energy 
$T$ of the system (in the centre-of-mass coordinate system) and its 
gravitational potential energy $U$ (taken positive) are: 
\begin{equation} 
	T = \frac{1}{2} \frac{m_{1}m_{2}}{M}V^{2} ~ , 
\end{equation} 
\begin{equation} 
	U = \frac{Gm_{1}m_{2}}{R} ~ , 
\end{equation} 
where $R$ is the size of the system and $G$ is the gravitational constant. 
These equations are related by a simple relation: 
\begin{equation} 
	U = 2T ~ . 
\end{equation}
However, the above relation holds only for isolated self-gravitating systems
when the system is in equilibrium. 

In general, groups of galaxies (and dark matter haloes\footnote{We will use the 
term 'halo' from now on to refer to virialized clumps of dark matter in the 
simulation and reserve 'galaxy' for the real observational data.}) contain more 
than two members. Therefore a generalized method for calculating the kinetic 
and the potential energies is needed. We adopt a method described by 
\citet{Virial}. In general, the kinetic energy may be written:
\begin{equation}
	T = \frac{1}{2M} \sum_{i < j} m_{i}m_{j}(\bmath{V}_{i} - \bmath{V}_{j} )^{2} ~ ,
\end{equation}
and the potential energy:
\begin{equation}
	U = G \sum_{i < j} \frac{m_{i}m_{j}}{R_{i,j}} ~ ,
\end{equation}
where $m_{i}$ and $m_{j}$ are the masses of the two galaxies, 
$\bmath{V}_{i}$ and $\bmath{V}_{j}$ are their velocities, and $R_{i,j}$ is the 
distance between them.

We use these general equations to get the total kinetic energy of a group of 
haloes and compare it to its total potential energy. If the group of haloes 
does not fulfill the criterion:
\begin{equation}
	T - U < 0 ~ ,
\end{equation}
it is entered into a list of unbound groups. The above criterion is equal to the virial
ratio:
\begin{equation}\label{virratio}
	\frac{T}{U} < 1.0 ~ ,
\end{equation}
which we use throughout this paper as a criterion for discriminating between
bound and unbound groups.

\section[]{Description of the cosmological simulations}\label{s4}

%This section contains a detailed description of the method used to set up our...

\subsection[]{Background}

We present results from four simulations, performed by the cosmological N-body 
simulation code AMIGA (Adaptive Mesh Investigations of Galaxy Assembly). The 
former version of AMIGA was known as MLAPM (for details see \citealt{AMIGA}). 
For the first two runs we adopt the currently popular flat low-density 
cosmological model $\Lambda$CDM with $h = 1.0$, $\Omega_{dm} = 0.27$, 
$\Omega_{\Lambda} = 0.73$ and $\sigma_{8} = 0.83$, with two different 
resolutions. Both simulations were made with $256^{3}$ dark matter particles. 
The high resolution simulation began at the initial redshift of $z_{i} = 47.96$ 
while the low-resolution simulation was initiated at redshift $z_{i} = 
38.71$. The volume employed in the high resolution simulation was 
$(40h^{-1}~\rmn{Mpc})^{3}$ and $(80 h^{-1}~\rmn{Mpc})^{3}$ in the low 
resolution simulation corresponding to the mass resolutions of $2.86 \times 
10^{8}h^{-1}~M_{\sun}$ and $2.29 \times 10^{9}h^{-1}~M_{\sun}$, respectively. 
The force resolution for the high resolution simulation is 
$1.8h^{-1}~\rmn{kpc}$ and for the low-resolution $7.3h^{-1}~\rmn{kpc}$.

For the third and the fourth simulations we adopt different cosmological 
models. These simulations were also performed with $256^{3}$ dark matter 
particles but with different values of the cosmological constant $\Lambda$. The 
total density of the universe was kept equal to the critical density ($\Omega = 
1.0$). For the third simulation we adopt  $h = 1.0$, $\Omega_{dm} = 0.1$, 
$\Omega_{\Lambda} = 0.9$ and $\sigma_{8} = 0.83$. For the fourth simulation we 
adopt $h = 1.0$, $\Omega_{dm} = 1.0$, $\Omega_{\Lambda} = 0.0$ and $\sigma_{8} 
= 0.84$ (see Table \ref{tb:reso}).

The high resolution $\Omega_{\Lambda} = 0.73$ simulation was used to understand 
the effects of limited resolution in N-body simulations. During this 
work, we found some differences between results of the high and low-resolution 
$\Omega_{\Lambda} = 0.73$ simulations. These differences are clearly visible
when the group abundances are studied (see Figures 1 $-$ 4).

\begin{table}
\caption{Details of the simulations analyzed in this paper.}
\label{tb:reso}
\begin{tabular}{ccccccc}
  \hline 
  $\Omega_{\Lambda}$ & $L$ & $N_{p}$ & $z_{i}$ & $m_{res}$ & $F_{res}$ & $N_{h}$ \\
  \hline 
  $0.73$ & $40$ & $256^{3}$ & $47.9600$ & $2.86\times10^{8}$ & $1.8$ & $6700$ \\
  $0.73$ & $80$ & $256^{3}$ & $38.7099$ & $2.29\times10^{9}$ & $7.3$ & $9301$ \\
  $0.90$ & $80$ & $256^{3}$ & $31.7858$ & $8.50\times10^{8}$ & $7.3$ & $4937$ \\
  $0.00$ & $80$ & $256^{3}$ & $72.8767$ & $8.47\times10^{9}$ & $3.7$ & $7919$ \\
  \hline
\end{tabular}

\medskip 
Note: {\em $\Omega_{\Lambda}$} specifies the value of the cosmological 
constant, $L$ is the size of the simulation box in one dimension in 
$h^{-1}\rmn{Mpc}$, $N_{p}$ is the number of dark matter particles, $z_{i}$ is 
the initial redshift, $m_{res}$ is the mass resolution in $h^{-1}M_{\odot}$, 
$F_{res}$ is the force resolution in $h^{-1}\rmn{kpc}$, and $N_{h}$ is the total 
number of dark matter haloes identified from the simulation.
\end{table}

\subsection[]{Halo finder and identification of the dark matter haloes}

Our simulations only follow the evolution of the dark matter particles via 
gravitational interaction. It is expected that baryons condense and form 
galaxies at the centres of dark matter haloes. AMIGA N-body simulation code 
comes with a halo finding algorithm called MHF (MLAPM's Halo Finder, 
\citealt{MHF}). For our purpose of analyzing the nearby groups of haloes we 
used MHF.

The general goal of a halo finder, such as MHF, is to identify gravitationally 
bound objects. MHF essentially uses the adaptive grids of the AMIGA to locate 
the haloes and the satellites of the host haloes, namely subhaloes. The 
advantage of reconstructing the grids to locate haloes is that they follow the 
density field with the exact accuracy of the simulation code and therefore no 
scaling length is required. For more detailed description of the MHF see 
\cite{MHF}.

The minimum number of particles in a halo was set to $10$. This corresponds to 
a halo mass $\sim 3\times10^{9}h^{-1}~M_{\odot}$ and $\sim 
2\times10^{10}h^{-1}~M_{\odot}$ for the high and the low-resolution simulation, 
respectively. A low value of the minimum number of particles in a halo ensures 
that even with a limited resolution, large and massive haloes are split into 
lighter subhaloes. For large and massive ($\sim 10^{14}h^{-1}~M_{\odot}$) 
haloes, subhaloes represent visual galaxies, masses $\sim 
10^{12}h^{-1}~M_{\odot}$. In a typical case, the total massfraction in 
subhaloes is $\sim 10$ per cent, and only these are visible in our 'mock' catalogue.
 In our 'mock' catalogue, the median of individual dark matter haloes mass is 
$\sim 1.6\times 10^{12}h^{-1}~M_{\odot}$ when the $\Omega_{\Lambda} = 0.73$ 
model and the low-resolution is adopted. The first and the third quartiles are: 
$\sim 7.2\times 10^{11}h^{-1}~M_{\odot}$ and $\sim 5.0 \times 
10^{12}h^{-1}~M_{\odot}$, respectively.

The AMIGA and its halo finder calculate automatically certain properties (e.g. 
position, mass, velocity etc.) of the dark matter haloes. These properties were 
used when the FOF algorithm was applied to generate the catalogues of groups of 
dark matter haloes. Subhaloes were included in our data as our purpose is to 
study if the groups of galaxies (dark matter haloes) are gravitationally bound 
objects. The results did not change substantially when the subhaloes of the 
more massive haloes were excluded from the analysis. This result is due to the 
small number of subhaloes in our low-resolution simulation. The results of the 
high resolution simulation show no significant difference if subhaloes were 
excluded because subhaloes have a relatively small mass. Due to their small masses, 
subhaloes are not visible at the observation point, when the apparent 
magnitude limit of $13.2$ is adopted.

\section[]{Statistical properties of the nearby groups of galaxies: comparing 
simulations with observations}\label{s5} 
\subsection[]{Selection of the nearby groups of haloes}

A Total of ten catalogues were generated, corresponding to 10 different 
'observers', for each simulation, with different apparent magnitude limits. 
Ten observers are used to produce enough groups to give good statistics. Note, 
however, that since all ten catalogues are constructed from the same parent 
simulation, the scatter between statistics estimated from them might 
underestimate the true sampling variance.

All observation points were chosen with the following criteria:
\begin{enumerate}
	\item observation point is $> 15h^{-1}$ Mpc from the edge of the simulation box
	\item a massive ($\sim 5 \times 10^{14}h^{-1}~M_{\odot}$) Virgo -type halo is 
	located within a distance of $\sim 20h^{-1}$ Mpc.
\end{enumerate}
We did not restrict the local $< 10h^{-1}$ Mpc environment of the observation 
points by any criteria. Although it is not clear if choosing an observation 
point simply by the two former criteria resolves the environment issue, we 
believe this to be strict enough for the statistical study of the virial ratio 
of groups. These criteria are justified for the statistical study, as we did not 
perceive a significant difference between the observation points in the low 
resolution simulation. Small differences between observation points were 
observed when the high-resolution simulation was studied, as it only includes two
Virgo -type haloes. When the low-resolution simulation was studied, the 
location of a massive (Virgo -type) halo did not have any significant effect to 
our results.

The simulation data do not directly give the luminosity or the absolute 
magnitude of the dark matter haloes, which are needed when we are mimicking 
observational conditions. We use halos' virial mass $M_{vir}$ to obtain its 
luminosity. To obtain the luminosity of an object in the blue band we use the 
relation proposed by \citet{Vale}: 
\begin{equation} 
	L(M_{vir}) = 5.7 \times 10^{9} h^{-2} L_{\odot} \frac{M_{11}^{p}}{\left[ q + 
	M_{11}^{s(p-r)} \right]^{1/s}} ~ , 
\end{equation} 
where $M_{11}$ is defined:
\begin{equation} 
	M_{11} = \frac{M_{vir}}{10^{11}h^{-1}M_{\odot}} ~ . 
\end{equation} 
For the free parameters of the mass-luminosity function, values of 
$p = 4.0$, $q = 0.57$, $r = 0.28$, and $s = 0.23$ were adopted 
(\citealt{Oguri}). It has been shown by \citet{Cooray} that the relation 
between the mass of a dark matter halo and its luminosity is not as 
straightforward as presented above. For our purposes, as the luminosity of a 
dark matter halo is used only to determine whether a halo is visible from the 
observation point, the above relation should be satisfactory. For this work we 
do not adopt more complex methods such as actual distributions for the 
mass-luminosity relation.

After the luminosity $L$ of the halo is known we obtain the apparent magnitude 
of the halo in the blue band from the equation:
\begin{equation}\label{Bmag}
	m_{B} = M_{\odot_{B}} -2.5 \log_{10}\left(\frac{L}{L_{\odot}}\right) + 5 
	\log_{10}\left(\frac{d}{1~\rmn{Mpc}}\right) + 25 ~ , 
\end{equation} 
where $d$ is the distance from the observation point and the magnitude of the 
sun in blue band $M_{\odot_{B}} = 5.47$ (\citealt{Allen}). As seen from
Equation \ref{Bmag}, we do not include dust extinction in our study, as its effect 
in a statistical study like ours would be negligible.

The method described above allows us to use the apparent magnitude limit 
$m_{lim} = 13.2$ as adopted in HG82. Group catalogues of this study are also 
generated with different magnitude limits in order to understand the effects of 
the magnitude limit in magnitude-limited samples. Unless explicitly noted, all 
haloes and groups referred to are from the simulations; real groups of galaxies 
from HG82 and UZC-SSRS2 (\citealt{RGPC}) are denoted as such.

\subsection[]{Comparison parameters}

%In this subsection we briefly present the equations and methods used in HG82 for %calculating parameters and properties of the groups of galaxies. In the next %subsection we make comparisons between simulations and observations by using the %equations given in this subsection.

We begin by calculating the velocity dispersion $\sigma_{v}$ of a group. In 
general, the velocity dispersion of a group is defined as: 
\begin{equation} 
	\sigma_{v} = \sqrt{ \frac{1}{N_{H} - 1} \sum_{i = 1}^{N_{H}}(v_{i} -
	<v_{R}>)^{2}} ~ , 
\end{equation} 
where $N_{H}$ is the number of haloes (or galaxies) in a group, $v_{i}$ is the 
radial velocity of the $i$th halo (or galaxy) and $<v_{R}>$ is the mean group radial 
velocity.

The second comparison parameter is the mean pairwise separation $R_{p}$ which 
is a measure of the size of a group. It can be defined as: 
\begin{equation} 
	R_{p} =  \frac{8<v_{R}>}{\pi H_{0}} \sin \left( \frac{1}{N_{H}(N_{H}-1)} 
	\sum_{j<i}^{N_{H}} \sum_{i=1}^{N_{H}} \theta_{ij} \right) ~ , 
\end{equation} 
where 
$<v_{R}>$ is the mean group radial velocity, $H_{0}$ is the Hubble constant, and 
$\theta_{ij}$ is the angular separation of the $i$th and $j$th group members. 
Other two comparison parameters are the total group mass and the virial
crossing time (in units of the Hubble time $H_{0}^{-1}$) which can be 
defined as:
\begin{equation} 
	t_{c} = \frac{3R_{H}}{5^{3/2}\sigma_{v}} ~ , 
\end{equation} 
where $\sigma_{v}$ is the velocity dispersion and $R_{H}$ is the mean harmonic 
radius: 
\begin{equation} 
	R_{H} = \frac{\pi <v_{R}>}{H_{0}} \sin \left\lbrace \frac{1}{2} 
	\left[ \frac{N_{H}(N_{H}-1)}{2} \left( \sum_{i=1}^{N_{H}} \sum_{j>i}^{N_{H}} 
	\theta_{ij} \right)^{-1} \right] \right\rbrace  %~ , 
\end{equation} 
where all the variables are defined as above.

In observations, the group masses can be estimated in various methods. In the 
HG82 and the UZC-SSRS2 catalogues the total mass of a group is estimated with a
simple relation:
\begin{equation}\label{obsmass}
	M_{obs} = 6.96\times10^{8}\sigma_{v}^{2}R_{H}M_{\odot} ~ ,
\end{equation}
where $\sigma_{v}$ is the velocity dispersion, and $R_{H}$ is the mean harmonic 
radius of the group as defined above. We use this simple relation to determine 
the 'observable' mass of a group when we are comparing the masses of the 
simulated groups to the real observed groups such as HG82 and UZC-SSRS2. 
However, when the total mass of a group is used to scale the properties of 
groups (in Section \ref{virial}), we calculate the total mass of a group of dark 
matter haloes as a sum of the member halo masses, namely the 'true' mass of the 
group. In case of the subhaloes, the diffuse dark matter is included in the 
main halo's mass. In general, we do not include the diffuse dark matter within a 
given distance from the group centre, as it would be troublesome to choose an 
appropriate distance. We do not consider this error to be meaningful, as the 
diffuse dark matter does not substantially give rise to the mean density of a 
simulation.

\subsection[]{Comparison with observations}

Comparison between simulations and observations are done using a 
Kolmogorov-Smirnov (K-S) test. The null hypothesis $H_{null}$ of the K-S test is 
that the two distributions are alike and are drawn from the same population 
distribution function. Results of the K-S tests are presented as significance 
levels (value of the $Q$ function) for the null hypothesis. Correlation between 
two variables is proved or disproved with the use of the linear correlation 
coefficient $r$. In general, the significance level of $0.001$ is adopted when 
the correlation between two variables is determined. For correlations we also 
present a probability $P(r)$ of observing a value of the correlation 
coefficient greater than $r$ for a sample of $N$ observations with $N-2$ 
degrees of freedom.

We begin the comparison of our simulations to observations by using the 
parameters presented in the previous subsection. A direct comparison between 
HG82 and simulations is possible as the magnitude limit ($13.2$) and the 
depth of the catalogues ($cz < 4,000$ km s$^{-1}$) are comparable. Comparisons 
with the more recent group catalogue UZC-SSRS2 is also done. The UZC-SSRS2 
catalogue has the magnitude limit of $15.5$ and only galaxies with $cz < 
15,000$ km s$^{-1}$ has been considered. These differences make the direct 
comparison of the UZC-SSRS2 with the simulations less conclusive. We also 
compare our ten observation points with each other but do not find significant 
difference between them. This justifies our method of choosing the 
observational points as stated before.

In Figures \ref{fig1} $-$ \ref{fig13} the abundance of groups is scaled to the 
volume of a sample as the distributions depend strongly on selection and 
volume effects. However, as there is no 'total' volume of a galaxy sample in 
magnitude-limited group catalogues, we weight each group according to its 
distance (\citealt{MFW} and \citealt{Diaferio}). As we only consider groups with 
three or more members, we can identify a group only when its 
third-brightest galaxy has an absolute magnitude
\begin{equation}
	M_{i} \leq m_{lim} - 25 - 5 \log(\frac{<cz>}{H_{0}}) ~ , 
\end{equation}
where $<cz>$ is the mean velocity of the group. $M_{i}$ 
determines the radius $cz_{i}$ of a sphere within which we could have 
identified this group. 

We calculate the comoving volume sampled by a group with the following equation:
\begin{equation}
	\Psi_{i} = \frac{\omega}{3}\left ( \frac{cz_{i}}{H_{0}} \right )^{3} \left [
	1- \frac{3z_{i}}{2} \left( 1 +  \frac{\Omega}{2} \right) \right] ~ ,
\end{equation}
where $\omega$ is the solid angle of the catalogue, $z_{i}$ is the redshift of 
the group, $c$ is the speed of light, and $\Omega$ is the cosmological density 
parameter, taken as $1.0$. Each group of galaxies (or dark matter haloes) 
contributes with a weight of $\Psi_{i}^{-1}$ to the total abundance of groups.
We include all galaxies with $cz > 500$ km s$^{-1}$. This lower cut-off avoids 
including faint objects that are close to the observation point as these 
groups could contain galaxies fainter than real magnitude-limited surveys. 
Therefore we consider only groups with $<cz>$ larger than $500$ km s$^{-1}$ in 
mock, HG82, and UZC-SSRS2 catalogues.

Figures \ref{fig1} $-$ \ref{fig4} show that the $\Lambda$CDM simulations are, 
in general, in a moderate agreement with observations when the resolution 
effects of the N-body simulations are taken into account. Our $\Lambda$CDM 
simulations are within 2$\sigma$ from the UZC-SSRS2 catalogue and within 
3$\sigma$ from the HG82 catalogue. Figures \ref{fig1} $-$ \ref{fig4} are all from 
the $\Omega_{\Lambda} = 0.73$ simulations when the apparent magnitude limit has 
been set to $13.2$, comparable to HG82. The error bars in Figures \ref{fig1} $-$ 
\ref{fig4} are the standard deviation between ten observation points. Unless 
explicitly noted, the simulations referred to are $\Omega_{\Lambda} = 0.73$ 
simulations, other models are denoted as such.

\subsubsection{Velocity dispersion}

\begin{figure}
\includegraphics[width=84mm]{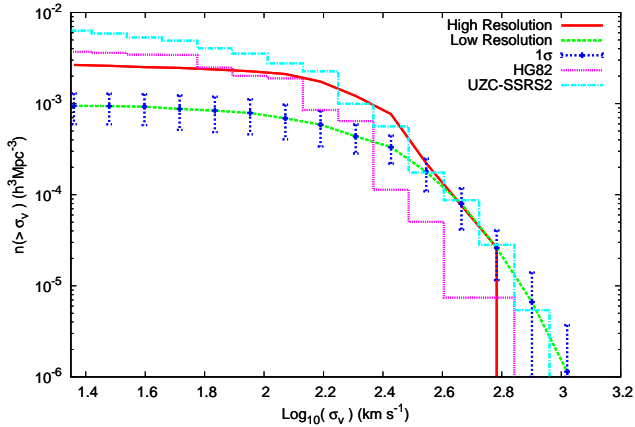}
\caption{The cumulative number density of velocity dispersion $\sigma_{v}$ for 
galaxy groups. Simulation data are from the $\Omega_{\Lambda} = 0.73$ 
simulations, and it is averaged over the ensemble of ten mock catalogues. The
error bars are 1$\sigma$ errors and are only shown for the low-resolution
$\Omega_{\Lambda} = 0.73$ simulation for clarity. The error bars for other data
have similar size.}
	\label{fig1}
\end{figure}

Figure \ref{fig1} shows that the cosmological $\Omega_{\Lambda} = 0.73$ model 
can produce velocity dispersions similar to observations (see also 
\citealt{Klypin}, \citealt{Maccio} and \citealt{PP}). Figure \ref{fig1} agrees 
with results by \cite{Luca} (their Figure 14) even though \cite{Luca} 
considered only groups with $> 5$ members. Our low-resolution simulation 
produces roughly the right number density of groups when the high ($> 100$ km 
s$^{-1}$) velocity dispersions are considered and the comparison is carried out 
against more recent observations (UZC-SSRS2). However, due to the limited mass 
resolution, the low-resolution simulation lacks a significant number of groups 
when the abundance of groups with velocity dispersions $< 100$ km s$^{-1}$ is 
studied. Because of this discrepancy the applied K-S test fails: $Q \sim 
10^{-6}$ (against the HG82) and $Q \sim 10^{-6}$ (against the UZC-SSRS2). 
Even-though the K-S test fails, the low-resolution simulation is 
within 3$\sigma$ from the HG82. When the high-resolution simulation is 
considered we get roughly the same number density of groups as in observations. 
However, the high resolution simulation lacks groups with velocity dispersions 
$> 500$ km s$^{-1}$. This can be explained by the small volume of the 
high-resolution simulation. Even with this discrepancy, the applied K-S tests 
are approved with the significance levels of $0.02$. When observations (HG82 
and UZC-SSRS2) are compared against each other, the applied K-S test is 
approved at level of $0.34$. For detailed significance levels of the K-S tests 
see Table \ref{tb:K-S}.

\subsubsection{Mass}

\begin{figure}
\includegraphics[width=84mm]{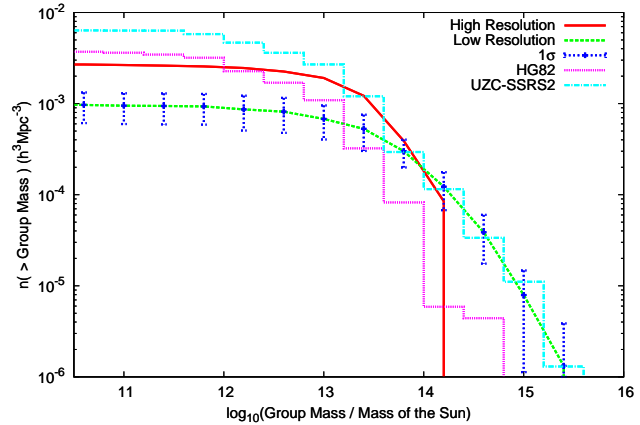}
\caption{Group abundance by 'observable' mass of the groups. Simulation data 
are from the $\Omega_{\Lambda} = 0.73$ simulations, and it is averaged over the 
ensemble of ten mock catalogues. The error bars are 1$\sigma$ errors and are 
only shown for the low-resolution $\Omega_{\Lambda} = 0.73$ simulation for 
clarity. The error bars for other data have similar size.}
	\label{fig2}
\end{figure}

Figure \ref{fig2} shows that the cosmological $\Lambda$CDM model can produce 
'observable' masses (Eq. \ref{obsmass}) similar to observations when the 
resolution effects of the simulations are considered. The low-resolution 
simulation can produce the same number density of groups as in the UZC-SSRS2 
catalogue when massive ($\log$(Group Mass M$_{\odot}^{-1}) > 13.5$) groups are 
considered. Both the UZC-SSRS2 catalogue and the low-resolution mock catalogue 
has an excess of massive groups if comparison is carried out against the HG82 
catalogue. When less massive groups ($\log$(Group Mass M$_{\odot}^{-1}) < 
13.0$) are studied, the low-resolution simulation has a number density of 
groups which is over 3$\sigma$ lower than in the UZC-SSRS2 catalogue. The 
resolution effect is clearly visible in Figure \ref{fig2} when the high 
resolution simulation is studied, as it can produce about the right number 
density of groups when less massive ($\log$(Group Mass M$_{\odot}^{-1}) < 
13.0$) groups are considered. The high-resolution simulation is less than 
1$\sigma$ away from the HG82 and within 2$\sigma$ from the UZC-SSRS2 even when 
groups with $\log$(Group Mass M$_{\odot}^{-1}) < 11.0$ are considered. The 
applied K-S test is approved ($Q \sim 0.58$) only when the high-resolution 
simulation is compared to the HG82 catalogue. In all other cases the K-S test 
fails. For numerical details of the K-S test, see Table \ref{tb:K-S}.

As the 'observable' mass of a group depends strongly on the groups' velocity 
dispersion (see Equation \ref{obsmass}) we made another comparison between 
group abundances by mass. If we use the 'true' mass of a group instead of an 
'observable' mass, other differences arise. There is no substantial difference 
between the plots of 'observable' and 'true' mass when comparing the abundances 
of lighter ($\log$(Group Mass M$_{\odot}^{-1}) < 14.0$) groups. However, our 
simulations do not produce a single group with a 'true' mass $> 
5\times10^{14}h^{-1}$M$_{\odot}$. The low and high-resolution $\Omega_{\Lambda} 
= 0.73$ simulations show the same cut-off, thus the lack of massive groups is 
not a resolution effect. However, the small volume of our simulation boxes 
explains the lack of massive groups in simulations and the existence of groups 
with large 'observed' mass is due to projection effects, which are not 
realiably taken into account in Eq. \ref{obsmass}.

\subsubsection{Size}

\begin{figure}
\includegraphics[width=84mm]{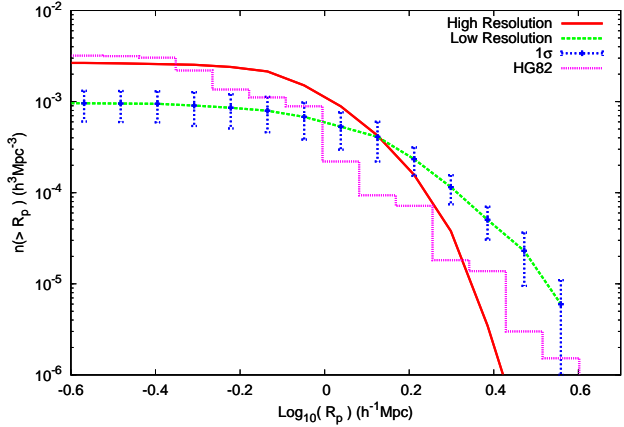}
\caption{The cumulative distribution of mean pairwise separation $R_{p}$ for 
galaxy groups. Simulation data are from the $\Omega_{\Lambda} = 0.73$ 
simulations, and it is averaged over the ensemble of ten mock catalogues. The 
error bars are 1$\sigma$ errors and are only shown for the low-resolution 
$\Omega_{\Lambda} = 0.73$ simulation for clarity. The error bars for other data 
have similar size.}
	\label{fig3}
\end{figure}

The cosmological $\Omega_{\Lambda} = 0.73$ model can produce groups of haloes 
which are similar in size to observed groups. Note, however, that we do not 
compare our simulations to the UZC-SSRS2 catalogue, as it doesn't contain the 
information about the pairwise separtion of groups. Our high-resolution 
simulation produces about the right number density of groups when small 
($\log$(R$_{p}$) $< -0.4$) groups are considered and the error is well within 
1$\sigma$. The high-resolution simulation seems to produce an excess of groups 
when intermediate size ($\log$(R$_{p}$) $\in [-0.3,0.4]$) groups are 
considered. However, as the only comparison observation is the HG82 catalogue 
this excess might not be as large as in Figure \ref{fig3}, as other comparisons 
(Figures \ref{fig1} and \ref{fig2}) show that the HG82 and the UZC-SSRS2 
catalogues differ quite significantly from each other.

When the low-resolution simulation is studied, we observe this same excess when 
larger ($\log$(R$_{p}$) $> 0.1$) groups are considered. Because of these 
discrepancies the applied K-S test fails in both cases, with $Q \sim 10^{-3}$ 
and $Q \sim 10^{-5}$ for the high and the low-resolution simulations, 
respectively. Even though the K-S test fails, the simulated mock catalogues of 
group abundances by mean pairwise separation are mostly within 2$\sigma$.

\subsubsection{Crossing time}

\begin{figure}
\includegraphics[width=84mm]{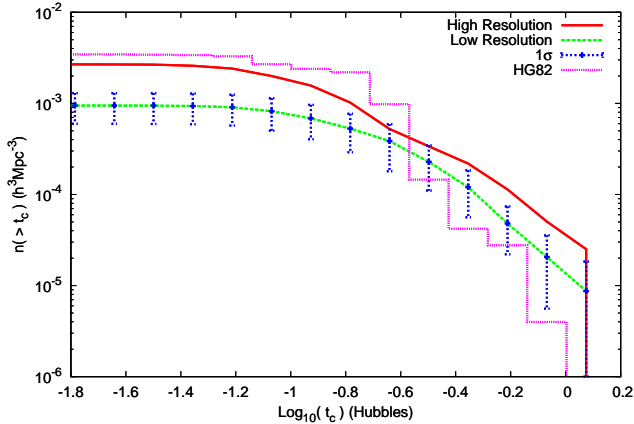}
\caption{The cumulative distribution of crossing time $t_{c}$ (in unit of the 
Hubble time) for galaxy groups. Simulation data are from the 
$\Omega_{\Lambda} = 0.73$ simulations, and it is averaged over the ensemble of 
ten mock catalogues. The error bars are 1$\sigma$ errors and are only shown for 
the low-resolution $\Omega_{\Lambda} = 0.73$ simulation for clarity. The error 
bars for other data have similar size.}
	\label{fig4}
\end{figure}

Figure \ref{fig4} shows that the cosmological $\Omega_{\Lambda} = 0.73$ model 
can produce groups with crossing times similar to the HG82 observations. The 
low-resolution simulation produces the number density of groups with small 
crossing times, which is a lot lower than observed. However, this discrepancy 
is due to limited resolution, as the high-resolution simulation produces a 
lot more groups with small crossing times. The high-resolution simulation 
produces roughly the right number density of groups when the crossing time of 
the group is studied. Some differences are observed when larger 
($\log$(t$_{c}$) $> -0.6$) crossing times are studied. Both simulations produce 
a higher number density than observed. For low-resolution simulation this 
excess is not large as the number density is within 2$\sigma$. Because of the 
discrepancies visible in Figure \ref{fig4}, the applied K-S test fails in both 
cases. For numerical details, see Table \ref{tb:K-S}. When more recent values 
of the Schechter luminosity function are adopted, somewhat lower crossing times 
are observed in general. However, more recent values of the Schechter 
luminosity function do not give a better agreement, and the K-S 
test fails. The significance levels of the K-S tests are $\sim 10^{-4}$ and 
$\sim 10^{-5}$, respectively.

Our simulations with the FOF algorithm do not (with a few 
exceptions) contain groups with crossing time larger than one Hubble time. For 
the high-resolution simulation, the median value of the crossing time is 
$0.14H_{0}^{-1}$. The median value of the crossing time for the HG82 catalogue 
and for the low-resolution simulation is $0.19H_{0}^{-1}$. Small crossing times 
suggests that groups of galaxies have had time to virialize and these groups 
should be gravitationally bound (see e.g. \citealt{GT}; \citealt{Tuck}; 
\citealt{Aceves} and \citealt{PBR}). We studied the correlation between 
crossing time and virial ratio and did not find any significant relation 
between these two variables. The linear correlation coefficient of $-0.01$ 
suggests that there is no correlation between crossing time and virial 
ratio, in the low-resolution simulation when the apparent magnitude limit of 
$13.2$ is adopted. This correlation is not significant at level of $0.05$ and 
$P(r) \sim 0.29$. There is no significant correlation between these two 
variables when different values of $\Omega_{\Lambda}$, resolutions or the 
apparent magnitude limits are adopted (see Table \ref{tb:crosst}). The lack of 
correlation between virial ratio and crossing time (see similar results in 
\citealt{DRGF}) calls into question the crossing time as an estimator of 
gravitationally bound systems which is widely accepted in observations.

\begin{table}
\caption{The correlation between crossing time and virial ratio when different values of 
$\Omega_{\Lambda}$ and the apparent magnitude limits are adopted.}
\label{tb:crosst}
\begin{tabular}{@{}lcrcc}
  \hline 
  $\Omega_{\Lambda}$ & $m_{lim}$ & $r$ & $\alpha$ & $P(r)$\\ 
  \hline 
  $0.00$	 & $13.2$ & $-0.009$  & $> 0.05$ & $0.33$\\
  $0.00$	 & $20.0$ & $ 0.005$  & $> 0.05$ & $0.36$\\
  $0.73^{H}$ & $13.2$ & $-0.016$  & $> 0.05$ & $0.26$\\
  $0.73^{L}$ & $13.2$ & $-0.016$  & $> 0.05$ & $0.29$\\
  $0.73^{L}$ & $20.0$ & $-0.016$  & $> 0.05$ & $0.09$\\
  $0.90$	 & $13.2$ & $-0.012$  & $> 0.05$ & $0.42$\\
  $0.90$	 & $20.0$ & $-0.022$  & $> 0.05$ & $0.09$\\
  \hline
\end{tabular}

\medskip 
Note: {\em $\Omega_{\Lambda}$} specifies the value of the cosmological constant 
($^{H}$ = high-resolution and $^{L}$ = low-resolution simulation), $m_{lim}$ is 
the apparent magnitude limit of the search, $r$ is the value of the linear 
correlation coefficient, $\alpha$ is the significance level, and $P(r)$ is the 
probability of observing a value of the correlation coefficient greater than 
$r$.
\end{table}

\subsubsection{Richness}

The $\Omega_{\Lambda} = 0.73$ model can produce groups comparable to 
observations when the number of members in a group is studied. The abundance of 
'rich' groups is roughly the same in the low-resolution simulation as in the 
observations. However, the low-resolution simulation cannot produce as many 
'poor' groups as is observed. The lack of 'poor' ($< 4$) and the excess of 
'intermediate' ($\in [6,40]$) groups is the reason why the K-S test fails ($Q 
\sim 10^{-5}$). The agreement is even worse ($Q \sim 10^{-7}$) when more recent 
values of Schechter luminosity function are adopted. These values ($\alpha = 
-1.15$, $M^{*} = -19.84$, and $\Phi^{*} = 0.0172$) produce a large number of 
'poor' groups and a lack of 'rich' groups. The difference between the 
low-resolution simulation and observations is due to the limited resolution in 
the simulations. When the high-resolution simulation is used, the agreement to 
observations, especially to UZC-SSRS2, is better ($Q \sim 10^{-3}$).

%\begin{figure}
%\includegraphics[width=84mm]{fig3.eps}
%\caption{Group abundance by 'observable' mass of the groups. The number of 
%groups is scaled to the 'volume' of the sample. Simulation data are from the 
%low resolution $\Omega_{\Lambda} = 0.73$ simulation when $m_{lim}=20.0$ is 
%adopted.
%	\label{fig5}
%\end{figure}

%\begin{figure}
%\includegraphics[width=84mm]{fig6.eps}
%\caption{The cumulative distribution of the number of galaxies in a group found 
%in the simulations and in the observations (HG82 and UZC-SSRS2). The number of 
%groups is scaled to the 'volume' of the sample. Simulation data are from the 
%$\Omega_{\Lambda} = 0.73$ simulation and it is averaged over the ensemble of 
%ten mock catalogues. The error bars are 1$\sigma$ errors and are only shown for 
%the low resolution $\Omega_{\Lambda} = 0.73$ simulation for clarity. The error 
%bars for other data have similar size.}
%	\label{nog}
%\end{figure}

\subsubsection{Influence of dark energy}

\begin{figure}
\includegraphics[width=84mm]{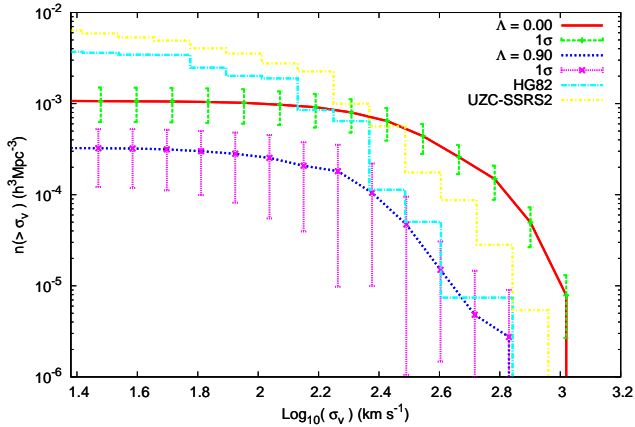}
\caption{The cumulative distribution of velocity dispersion $\sigma_{v}$ for 
galaxy groups from two different simulations with two different values of the 
cosmological constant $\Omega_{\Lambda}$ when the apparent magnitude limit of 
$13.2$ has been adopted. The simulation data are averaged over the 
ensemble of ten mock catalogues. For comparison also the observational data 
from the HG82 and the UZC-SSRS2 catalogue have been plotted. The standard 
$\Omega_{\Lambda} = 0.73$ simulations are shown in Figure \ref{fig1}.}
	\label{fig13}
\end{figure}

There are big differences between different cosmological models when dynamical 
properties of groups of dark matter haloes are studied. Figure \ref{fig13} 
shows the impact of dark energy on the formation of galaxy groups. It is 
clear that the $\Omega_{\Lambda} = 0.0$ simulation over produces groups 
with high velocity dispersions. The excess is over 3$\sigma$ if the comparison 
is carried out to the HG82 catalogue. A smaller discrepancy is observed when the
 comparison is carried out to the UZC-SSRS2 catalogue. The small number density 
of small velocity dispersion groups can be explained by the resolution effect 
which is also visible in Figure \ref{fig1}. Because of these discrepancies the 
K-S tests fails. When the $\Omega_{\Lambda} = 0.90$ simulation is studied, a 
qualitatively better agreement is observed, especially when the comparison is carried 
out to the HG82 catalogue. The great difference in the number density of small 
velocity dispersion groups can be explained by the resolution effect and the 
fact that the $\Omega_{\Lambda} = 0.90$ simulation has relatively small number 
of groups. Because of the great discrepancy in the number density of small 
velocity dispersion groups, the applied K-S test fails in both cases.

The $\Omega_{\Lambda} = 0.0$ cosmology produces more massive groups with 
greater velocity dispersion than the $\Lambda$CDM cosmology. Larger number of 
massive groups can partially be explained by the somewhat lower mass resolution 
in the $\Omega_{\Lambda} = 0.0$ simulation. However, the excess of massive 
groups is most likely due to Equation \ref{obsmass}, which we use to obtain the 
'observable' mass of a group, which depends strongly on the velocity dispersion 
of the group. For numerical details of K-S tests, see Table \ref{tb:K-S}.

\begin{table}
\caption{Comparison of HG82, UZC-SSRS2, and simulations when the 
apparent magnitude limit of $13.2$ is adopted.}
\label{tb:K-S}
\begin{tabular}{llcc|c}
  \hline 
  $\Omega_{\Lambda}$	&	& HG82 & UZC-SSRS2 & HG82 vs. UZC-SSRS2 \\
  \hline
  $0.00$	& $\sigma_{v}$	& $10^{-5}$ & $10^{-6}$ & $0.34$ \\ 
  $0.73^{H}$& $\sigma_{v}$ 	& $0.02$ 	& $0.02$ & $0.34$\\
  $0.73^{L}$& $\sigma_{v}$ 	& $10^{-6}$ & $10^{-6}$ & $0.34$\\ 
  $0.90$	& $\sigma_{v}$ 	& $10^{-6}$ & $10^{-7}$ & $0.01$\\   
  $0.00$	& $M_{G}$		& $10^{-5}$ & $10^{-6}$ & $ 10^{-3}$\\
  $0.73^{H}$& $M_{G}$		& $0.58$ 	& $10^{-5}$ & $ 10^{-3}$\\
  $0.73^{L}$& $M_{G}$		& $10^{-6}$ & $10^{-6}$ & $ 10^{-3}$\\
  $0.90$	& $M_{G}$		& $10^{-6}$ & $10^{-7}$ & $ 10^{-3}$\\	  
  $0.73^{H}$& $R_{p}$		& $10^{-3}$ & $-$ & $-$\\
  $0.73^{L}$& $R_{p}$		& $10^{-5}$ & $-$ & $-$\\ 
  $0.73^{H}$& $t_{c}$		& $10^{-4}$ & $-$ & $-$\\
  $0.73^{L}$& $t_{c}$		& $10^{-5}$ & $-$ & $-$\\
  \hline
\end{tabular}

\medskip 
Note: Significance levels of the K-S test for the null hypothesis that 
observations and the simulations ($^{H}$ = high-resolution and $^{L}$ = low 
resolution) are alike and are drawn from the same parent population (HG82 and 
UZC-SSRS2 columns). Significance levels of the K-S test for the null hypothesis 
that the HG82 and the UZC-SSRS2 group catalogue are alike and are drawn from 
the same parent population (HG82 vs. UZC-SSRS2 column).
\end{table}

\subsubsection{Median values and other properties}

The median values of the group properties are presented in Table 
\ref{tb:fractiles}. In general, our simulations seem to produce groups which 
median value of the velocity dispersion and the group mass is greater than in 
observations. In simulations, groups have also a greater median value for the
mean pairwise separation than in the HG82 sample. The $\Omega_{\Lambda} = 0.73$ 
simulations have the median value of velocity dispersions which are close to 
observations, even though they are somewhat higher. In general, median 
values of the group properties are in a moderate agreement with the results of 
similar studies (e.g. \citealt{Luca} and \citealt{Diaferio}). \cite{Luca} found 
larger values for the median velocity dispersions and group masses, but they 
considered only groups with $> 5$ members, which most likely makes the median 
values of groups somewhat higher.

\begin{table*}
\begin{minipage}{150mm}
\caption{Weighted quartiles of the group properties.}
\label{tb:fractiles}
\begin{tabular}{lcccccc}
  \hline 
  & $\Omega_{\Lambda}=0.00$ & $\Omega_{\Lambda}=0.73^{H}$ & $\Omega_{\Lambda}=0.73^{L}$ &
  $\Omega_{\Lambda}=0.90$ & HG82 & UZC-SSRS2\\ 
  \hline 
  $\sigma_{v}$ & $178/307/403$ & $135/180/295$ & $103/160/298$ & $124/209/271$ &
  $60/135/155$ & $60/130/178$ \\ 
  $\log M_{G}$  & $13.6/14.0/14.4$ & $12.7/13.2/13.5$ & $12.8/13.6/14.0$ &
  $12.7/13.1/13.5$ & $12.0/12.4/13.2$ & $12.1/12.8/13.3$ \\
  $R_{p}$ & $0.85/1.38/1.57$ & $0.81/1.21/1.47$ & $0.80/0.99/1.24$ & $0.72/1.01/1.20$ 
  & $0.44/0.54/1.00$ & $-$\\
  $t_{c}$ & $0.09/0.14/0.25$ & $0.07/0.14/0.19$ & $0.10/0.19/0.27$ 
  & $0.09/0.17/0.23$ & $0.10/0.19/0.27$ & $-$\\
  \hline
\end{tabular}

\medskip 
Note: Quantities $\sigma_{v}$, $\log M_{G}$, $R_{p}$, and $t_{c}$ are in units
of km s$^{-1}$, h$^{-1}$M$_{\odot}$, h$^{-1}$ Mpc, and Hubbles, respectively.

\end{minipage}
\end{table*}

The fractions of isolated galaxies, binary galaxies and groups of galaxies were 
also studied. If we compare our results with HG82, a significant difference in 
the fraction of isolated galaxies is noticed. Comparison to the LEDA 
(\citealt{Giudice}) catalogue shows a better fit. For more details, see Table 
\ref{tb:fractions}. \citet{Tuck} listed a large fraction of galaxies in 
groups from different group catalogues. These results and comparisons are not 
shown in this paper due to the different magnitude limits and grouping 
algorithms adopted in those observations. We may state that, in general, 
simulations show similar results to observations (excluding HG82), with regard to
the fractions of groups, binaries and isolated galaxies.

\begin{table}
\caption{Fractions of isolated galaxies, binary galaxies, and groups.}
\label{tb:fractions}
\begin{tabular}{ccccc}
  \hline Source & $m_{lim}$ & Groups ($\%$)  & Binaries ($\%$) & Isolated ($\%$) \\
  \hline 
  Simu & $13.2$ & $38.1\pm0.5$ & $19.7\pm0.1$ & $42.2\pm0.5$ \\
  Simu & $20.0$ & $41.2\pm0.2$ & $17.6\pm0.1$ & $41.2\pm0.2$ \\
  HG82 & $13.2$ & $60.0$ & $14.0$ & $26.0$ \\
  LEDA & $14.0$ & $33.3\pm0.5$ & $17.1\pm0.3$ & $49.6\pm0.4$ \\
  \hline
\end{tabular}

\medskip 
Note: Source refers to the sample (Simu $=$ the low-resolution 
$\Omega_{\Lambda} = 0.73$ simulation), $m_{lim}$ is the apparent magnitude 
limit (in $m_{B(0)}$ except for LEDA in $B_{T}^{0}$), Groups ($\%$) is the 
fraction of galaxies in the groups, Binaries ($\%$) is the fraction of galaxies 
forming double systems, and Isolated ($\%$) is the fraction of galaxies which 
are not classified into any group or double system. Poisson error limits have 
been calculated for the samples.
\end{table}

We also made an attempt to study discordant redshifts in compact groups 
observed e.g. by \citet{Sulentic} and \citet{Girardi}. This effect has been 
studied by several authors (see e.g. \citealt{Byrd}, \citealt{Mauri} and 
\citealt{Iovino}) who have come up with different explanations. According to 
these authors, apparent discordant redshifts arise when groups are not 
virialized and their central galaxies are incorrectly identified. Our findings 
are not conclusive as we did not have any exact method to identify which dark 
matter haloes might represent observable spiral galaxies. We did not 
observe any significant asymmetry in the radial velocities of the groups and 
neither this asymmetry was seen in the groups, which were misidentified (so that 
the brightest member is not the dominant member). No significant difference for 
the radial velocity asymmetry was discovered between bound and unbound groups.

\section[]{Gravitationally bound groups}\label{virial}

Gravitationally bound groups are determined by using the criterion (virial 
ratio, Eq. \ref{virratio}) presented in Section \ref{s3}. This method of computing the 
gravitational potential well of a group does assume that the group is isolated. 
This is not strictly true as each group is embedded in the large-scale matter 
distribution, which might have an effect to the threshold $1.0$ of the virial 
ration $T/U$. However, we believe this effect to be negligible in a statistical
study like ours.

Our study shows that $\sim 20$ per cent of groups generated by the FOF 
algorithm are not gravitationally bound when the $\Omega_{\Lambda} = 0.73$ 
model is adopted. This result is in agreement with \cite{D1994}, who derived a
similar result for the compact groups of galaxies. If we vary the apparent 
magnitude limit of the search from the original $13.2$ to $20.0$, even more 
groups ($\sim 37$ per cent) are unbound. This is not a negligible fraction 
considering that one widely accepted and applied method of calculating a group 
mass, from observations, is based on the assumption that groups found by the FOF 
algorithm are, in general, gravitationally bound systems.

If we vary the value of the cosmological constant from the original $0.73$ to 
$0.90$, a slightly larger fraction of groups seems to be unbound when the 
apparent magnitude limit of $13.2$ is adopted. This result is intuitively 
reasonable. If the negative vacuum pressure of space is larger, gravitational 
force becomes 'weaker' and a smaller number of dark matter haloes are formed and 
fewer groups are gravitationally bound objects. How does the fraction of 
gravitationally unbound groups change when the negative vacuum pressure of 
space is lowered? If the value of the cosmological constant is put to 
$0.0$, about the same fraction of groups (with $m_{lim} = 13.2$) are spurious 
as in the $\Omega_{\Lambda} = 0.90$ cosmology. When the apparent magnitude 
limit is changed to $20.0$, $\sim 37$ per cent of the groups are spurious (for 
details, see Table \ref{tb:virialization}).

\begin{table}
\caption{Fractions of gravitationally bound groups of dark matter haloes when
different cosmological models and apparent magnitude limits have been adopted.}
\label{tb:virialization}
\begin{tabular}{lcccc}
  \hline 
  $\Omega_{\Lambda}$ & $m_{lim}$ & $N_{groups}$  & $f_{bound}$ ($\%$) &
  $N_{isolated}$ ($\%$) \\ 
  \hline 
  $0.00$	 & $13.2$ & $2675$  & $79.2$ & $20.9$\\
  $0.00$	 & $20.0$ & $6213$  & $64.5$ & $41.8$\\
  $0.73^{H}$ & $13.2$ & $1570$	& $70.1$ & $32.8$\\
  $0.73^{L}$ & $13.2$ & $1168$  & $81.1$ & $42.2$\\
  $0.73^{L}$ & $20.0$ & $6807$  & $62.7$ & $41.2$\\
  $0.90$	 & $13.2$ & $238$   & $77.3$ & $53.1$\\
  $0.90$	 & $20.0$ & $3678$  & $62.0$ & $36.5$\\
  \hline
\end{tabular}

\medskip 
Note: {\em $\Omega$$_{\Lambda}$} specifies the value of the cosmological 
constant ($^{H}$ = high-resolution and $^{L}$ = low-resolution simulation), 
$m_{lim}$ is the apparent magnitude limit of the search, $N_{groups}$ is the 
number of groups found from ten observation points, $f_{bound}$ is the fraction 
of gravitationally bound groups, and $N_{isolated}$ is the percentage of the 
isolated haloes which do not belong to any group or binary system.
\end{table}

In the low-resolution $\Omega_{\Lambda} = 0.73$ simulation the fraction of 
gravitationally bound groups rises from $81.1$ to $81.7$ per cent, when more 
recent values of the Schechter luminosity function are adopted. Meanwhile the 
total number of groups decreases $\sim 10.0$ per cent. The fraction of 
gravitationally bound groups rises from $81.1$ to $82.8$ per cent, when the 
values of the free parameters of $D_{0} = 0.37~\rmn{Mpc}$ and $V_{0} = 
200~\rmn{km}~\rmn{s}^{-1}$ are adopted. This result agrees with Frederic 
(1995a,b) who obtained similar result while studying group accuracy as a 
function of $D_{0}$ and $V_{0}$. Frederic (1995a,b) showed that smaller values 
of $D_{0}$ and $V_{0}$ produces groups with greater accuracy and these groups 
should be gravitationally bound.

Adopting the values of $D_{0} = 0.37~\rmn{Mpc}$ and $V_{0} = 
200~\rmn{km}~\rmn{s}^{-1}$ have a significant effect to the total number of 
groups found from the simulations. The low-resolution $\Omega_{\Lambda} = 0.73$ 
simulation produces, in all, $1168$ groups of dark matter haloes when the 
original values ($D_{0} = 0.63~\rmn{Mpc}$ and $V_{0} = 
400~\rmn{km}~\rmn{s}^{-1}$) of the free parameters are adopted. When $D_{0} = 
0.37~\rmn{Mpc}$ and $V_{0} = 200~\rmn{km}~\rmn{s}^{-1}$ are adopted, the total 
number of groups found from the low-resolution simulation drops to $661$ while 
the fraction of isolated haloes rises from $\sim 42.2$ to $\sim 64.9$ per cent. 
Also, the group abundances changes significantly as the richest group found from 
the low-resolution simulation with $D_{0} = 0.37~\rmn{Mpc}$ and $V_{0} = 
200~\rmn{km}~\rmn{s}^{-1}$ has only $15$ members. These results are due to the 
fact that the limiting density enhancement of the search is inversely 
proportional to $D_{0}^{3}$.

Our study shows that the $\Omega_{\Lambda} = 0.0$ model produces about the same 
fraction of bound groups as the $\Omega_{\Lambda} = 0.90$ model when the 
apparent magnitude limit is $13.2$. However, the $\Omega_{\Lambda} = 0.73$ 
model produces more gravitationally bound systems than the two other models we 
study when the original apparent magnitude limit is adopted. If the apparent 
magnitude limit is changed to $20.0$, the $\Omega_{\Lambda} = 0.0$ model 
produces a slightly larger fraction of gravitationally bound groups than the 
$\Omega_{\Lambda} = 0.73$ or the $\Omega_{\Lambda} = 0.90$ simulations (for 
details see Table \ref{tb:virialization}). The use of the apparent magnitude 
limit of $20.0$ means simply that every single dark matter halo in a simulation 
box is visible at the observation point. This result is not without bias as the 
simulation box is of finite size and the edge effects might become significant, 
even using the periodic boundary conditions in the simulations.

\begin{figure}
\includegraphics[width=84mm]{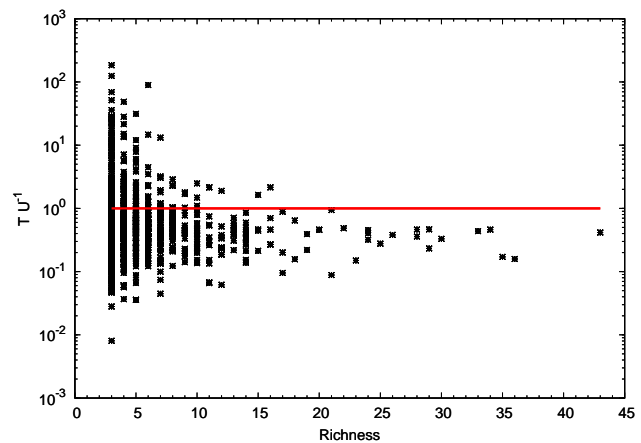}
	\caption{Virial ratio ($T U^{-1}$) versus the number of members in a group
	(Richness). Groups with more than 10 members are more often bound than 'poor' 
	groups with three to five members. The data are from the low-resolution 
	$\Omega_{\Lambda} = 0.73$ simulation when the apparent magnitude limit of $
	13.2$ has been adopted.}
	\label{fig6}
\end{figure}

The calculation that determines whether a group is bound is based on three 
parameters: the total mass of the group, the relative velocity of the group 
members and the physical size of the group. In the following we will study how 
sensitive the result is on the values of these parameters. But first we will 
study the virial ratio as a function of the number of members in the group. 
Figure \ref{fig6} shows the virial ratio $T/U$ as a function of the number of 
haloes in the group, namely richness. The data come from the low-resolution 
$\Omega_{\Lambda} = 0.73$ simulation with $m_{lim}=13.2$. There are, in all, 
$1168$ groups of haloes seen from ten different observation points. The 
fractions of bound 'poor' and 'rich' groups are shown in Table 
\ref{tb:halodiffvir}.

From Figure \ref{fig6} we see that groups with more than 10 members are most 
likely gravitationally bound and groups with three to five members are quite 
often unbound. \citet{RPG} argues that among groups with three members, $50$ to 
$75$ per cent of groups are spurious. They also conclude that for groups with 
more than three members the fraction of spurious groups is less than $30$ per 
cent and may be as small as $10$ per cent. Our findings are similar, and the fraction
 of bound groups with four or more members is about as high as \citet{RPG} suggested. 
\citet{RGPC} find that for groups with five or more members at least $80$ per 
cent of the groups are probably physical systems, but that $40$ to $80$ per 
cent of the groups with five or more members are bound groups. Our findings 
confirm the latter result. However, these results cannot be directly compared 
with ours, as slightly different values of the free parameters are adopted for 
the FOF algorithm. Even-though the free parameters of the FOF seem to have only 
a small effect to the fraction of spurious groups in our study.

Our findings for the $\Omega_{\Lambda} = 0.73$ cosmology are similar to 
\citet{RPG} for 'poor' (three or four members) groups, as can be seen in Figure 
\ref{fig6} and Table \ref{tb:halodiffvir}. However, our findings do not confirm 
the claim by \citet{RPG} that among groups with three members, $50$ to $75$ per 
cent of groups are spurious as we find only $\sim 23$ per cent of groups with three 
members to be gravitationally unbound with $m_{lim} = 13.2$. For the apparent 
magnitude-limited sample ($m_{lim} = 13.2$ comparable to HG82), we found  $\sim 
21$ per cent of the groups with three or four haloes to be spurious. 'Rich' 
groups with five or more members are more often gravitationally bound than 
'poor' groups, but the difference is relatively small at the apparent magnitude 
limit of $13.2$ as for 'rich' groups, we found $\sim 15$ per cent of the groups 
to be spurious. This is close to the upper limit proposed by Ramella et al. 
(1997, 2002). More details of our findings with different abundances and 
apparent magnitude limits are listed in Table \ref{tb:halodiffvir}.

\begin{table}
\caption{Fractions of gravitationally bound 'poor' and 'rich' groups of dark 
matter haloes generated from the low-resolution $\Omega_{\Lambda} = 0.73$ simulation.}
\label{tb:halodiffvir}
\begin{tabular}{cccc}
  \hline 
  $m_{lim}$ & $N_{haloes}$ & $N_{groups}$ & $f_{bound}$ ($\%$) \\
  \hline 
  $13.2$ & $3$ & $448$ & $77.0$ \\
  $13.2$ & $ \in [3,4]$ & $697$ & $78.5$ \\
  $13.2$ & $> 4$ & $471$ & $84.9$ \\
  $20.0$ & $3$ & $2786$ & $55.6$ \\
  $20.0$ & $ \in [3,4]$ & $4147$ & $56.3$ \\
  $20.0$ & $> 4$ & $2660$ & $72.6$ \\
  \hline
\end{tabular}

\medskip 
Note: {\em $m$$_{lim}$} is the apparent magnitude limit of a sample, 
$N_{haloes}$ is the number of haloes in a group, $N_{groups}$ is the number of 
groups found from ten observation points with appropriate number of haloes, 
and $f_{bound}$ is the fraction of gravitationally bound groups.
\end{table}

\begin{figure}
\includegraphics[width=84mm]{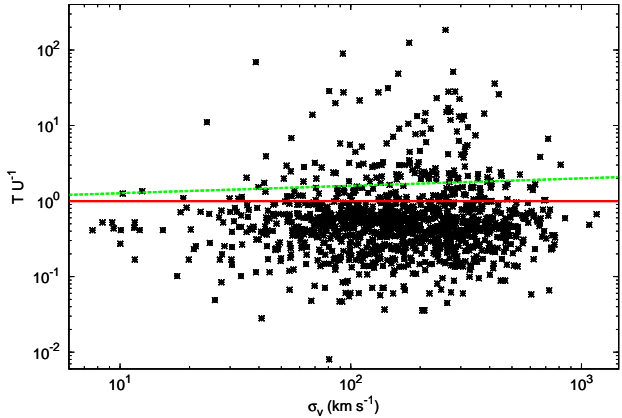}
\caption{Virial ratio ($T U^{-1}$) versus the velocity dispersion 
($\sigma_{v}$) of a group (in km s$^{-1}$). The rms straight line has been 
fitted to the data. The data are from the low-resolution $\Omega_{\Lambda} = 
0.73$ simulation when the apparent magnitude limit of $13.2$ has been adopted.}
	\label{fig7}
\end{figure}

\begin{figure}
\includegraphics[width=84mm]{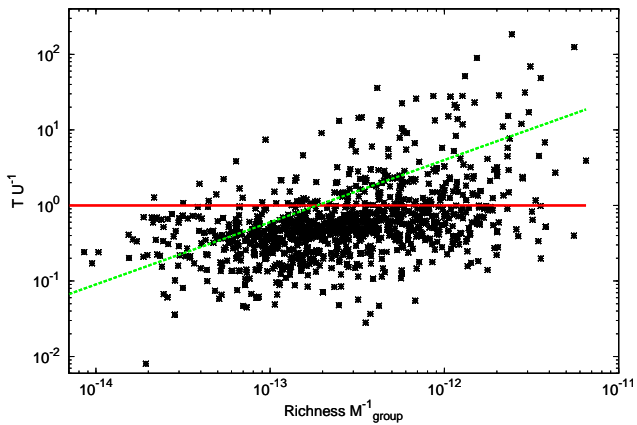}
	\caption{Virial ratio ($T U^{-1}$) versus the number of members (Richness) in a
	group when the latter has been scaled with the total mass of the group 
	(M$^{-1}_{group}$ in $h^{-1}~M_{\odot}$). The rms straight line has been fitted 
	to the data. The data are from the low-resolution $\Omega_{\Lambda} = 0.73$ 
	simulation when the apparent magnitude limit of $13.2$ has been adopted.}
	\label{fig8}
\end{figure}

\begin{figure}
\includegraphics[width=84mm]{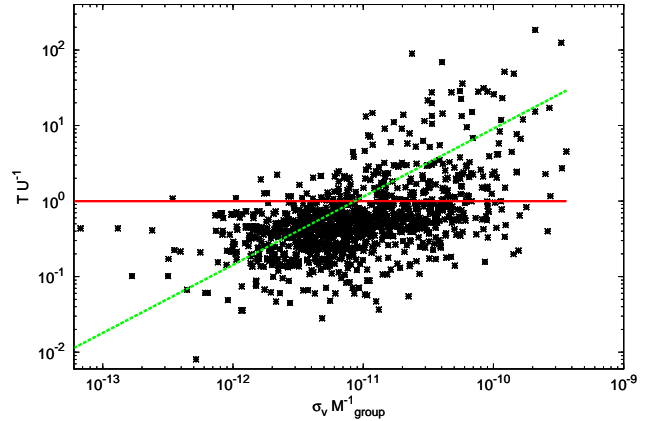}
\caption{Virial ratio ($T U^{-1}$) versus the velocity dispersion ($\sigma_{v}$) of a group 
(in km s$^{-1}$) when the latter has been scaled with the total mass of the 
group (M$^{-1}_{group}$ in $h^{-1}~M_{\odot}$).The rms straight line has been 
fitted to the data. The data are from the low-resolution $\Omega_{\Lambda} = 
0.73$ simulation, when the apparent magnitude limit of $13.2$ has been adopted.}
	\label{fig9}
\end{figure}

\begin{figure}
\includegraphics[width=84mm]{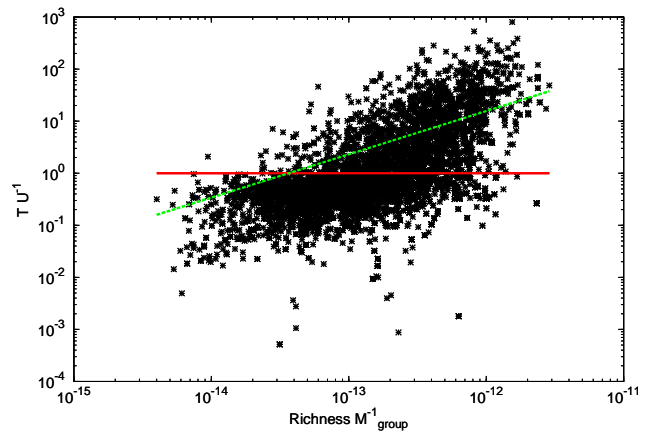}
	\caption{Virial ratio ($T U^{-1}$) versus the number of members in a group
	(Richness) when the latter has been scaled with the total mass of a group 
	(M$^{-1}_{group}$ in $h^{-1}~M_{\odot}$), and $\Omega_{\Lambda} = 0.0$ and 
	$m_{lim}=20.0$ has been adopted. The rms straight line has been fitted to the data.}
	\label{fig10}
\end{figure}

\begin{figure}
\includegraphics[width=84mm]{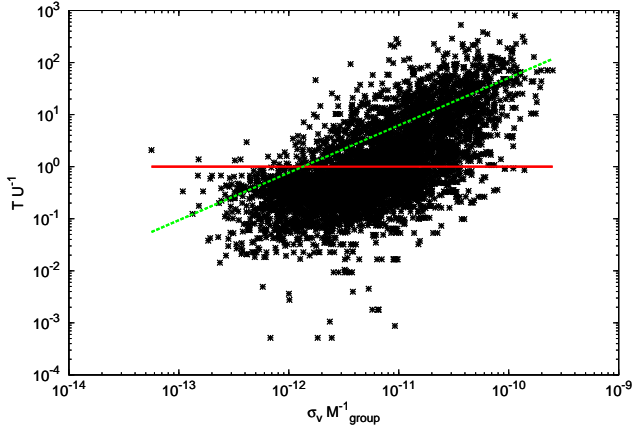}
	\caption{Virial ratio ($T U^{-1}$) versus the velocity dispersion ($\sigma_{v}$) of a group
	(in km s$^{-1}$) when the latter has been scaled with the total mass of a group 
	(M$^{-1}_{group}$ in $h^{-1}~M_{\odot}$), and $\Omega_{\Lambda} = 0.0$ and 
	$m_{lim}=20.0$ has been adopted. Straight line is a rms fit to the data.}
	\label{fig11}
\end{figure}

In Figure \ref{fig7}, the virial ratio is plotted as a function of the 
velocity dispersion $\sigma_{v}$ of the group. The plot shows a weak 
correlation in the sense that groups with large velocity dispersion are more 
often gravitationally unbound than groups with small velocity dispersion. The 
linear correlation coefficient of $0.17$ suggests that the correlation in 
Figure \ref{fig7} is weak. However, the correlation is significant at level of 
$0.001$ and $P(r) \sim 10^{-8}$. The rms line plotted in Figures \ref{fig7} $-$ 
\ref{fig11} is of the form $\frac{T}{U} \propto \sigma_{v}^{b}$ or $\frac{T}{U} 
\propto N(haloes)^{b}$. The value of the parameter $b$ of the rms line in 
Figure \ref{fig7} is $b = 0.10 \pm 0.04$.

The weak trends are clearer if we scale the abscissa in both Figures 
\ref{fig6} and \ref{fig7} with the total mass of the group. Note that we use 
here the 'true' mass of a group rather than the 'observable' mass. Results are 
shown in Figures \ref{fig8} and \ref{fig9}. More significant trends are now 
seen in both Figures, even-though the data are still scattered. The linear 
correlation coefficient of $0.32$ suggests that a significant correlation 
exists between the number of haloes and the virial ratio when the first is 
scaled with the total mass of the group. The correlation in Figure \ref{fig8} 
is significant at level of $0.001$ and $P(r) < 10^{-25}$. The slope of the rms 
line in Figure \ref{fig8} is $b = 0.82 \pm 0.03$.

Figure \ref{fig9} shows a strong trend even-though the data are still quite 
scattered. The linear correlation coefficient of $0.40$ 
suggests that the correlation between the velocity dispersion of a group and 
the virial ratio of a group is quite strong. The correlation in Figure 
\ref{fig9} is significant at level of $0.001$ and $P(r) < 10^{-25}$. The slope with
 standard errors of the rms line is now $b = 0.90 \pm 0.03$.

When the apparent magnitude limit is changed to $20.0$, the trends of Figures 
\ref{fig8} and \ref{fig9} become stronger and the asymptotic standard errors 
for the rms lines become much smaller. The linear correlation coefficient of 
$0.62$ shows that the correlation between the velocity dispersion of a group,
and the virial ratio, is strong when the apparent magnitude limit of $20.0$ is 
adopted. This correlation is significant at level of $0.001$ and $P(r) < 
10^{-25}$. What might be surprising is that changing the cosmological model,
i.e. the value of the cosmological constant $\Omega_{\Lambda}$, does not have a 
substantial influence on Figures 9 and 10. The number of groups found from 
different simulations varies a lot as a function of $\Omega_{\Lambda}$ but the 
fraction of gravitationally bound groups do not (see Table
\ref{tb:virialization}). For comparison with Figures \ref{fig8} and \ref{fig9}, 
we show results from the $\Omega_{\Lambda} = 0.0$ simulation in Figures 
\ref{fig10} and \ref{fig11}. In these figures the apparent magnitude limit of 
$20.0$ has been adopted. Otherwise, Figures \ref{fig10} and \ref{fig11} are 
comparable to Figures \ref{fig8} and \ref{fig9}. The linear correlation 
coefficient in Figures \ref{fig10} and \ref{fig11} is: $0.32$ and $0.65$, 
respectively. Both of the correlations are significant at a level of $0.001$ and
 $P(r) < 10^{-25}$ for both samples. The asymptotic standard errors for the rms 
lines in Figures \ref{fig10} and \ref{fig11} are small: $b = 0.83 \pm 0.02$ and 
$b = 0.91 \pm 0.01$, respectively.

\section{Discussion}\label{discussion}
\subsection{Probability functions of unbound groups}

In this subsection we briefly discuss a method, which gives a theoretical 
probability of a group being gravitationally unbound. The mass of the groups is 
assumed to be known. In observations, estimations of group masses are less than 
precise at best, therefore the applicability of this method to observational 
data is merely hypothetical. The observable quantities we study are the 
velocity dispersion divided by the group mass $\sigma_{v}M_{group}^{-1}$ and 
the mean pairwise separation divided by the group mass $R_{P}M_{group}^{-1}$.

To calculate the probability functions for the groups, the first step is to 
choose an appropriate bin length (generally between $0.15$ and $0.30$) in the 
logarithm of the observable quantity. Then one calculates the number of groups 
and the number of gravitationally bound groups in each bin and divides the 
number of unbound groups with the total number of groups in the bin. The 
logarithmic scale is chosen in order to lower the dispersion of the data and to 
assure a large enough number of groups in every bin.

\begin{figure}
\includegraphics[width=84mm]{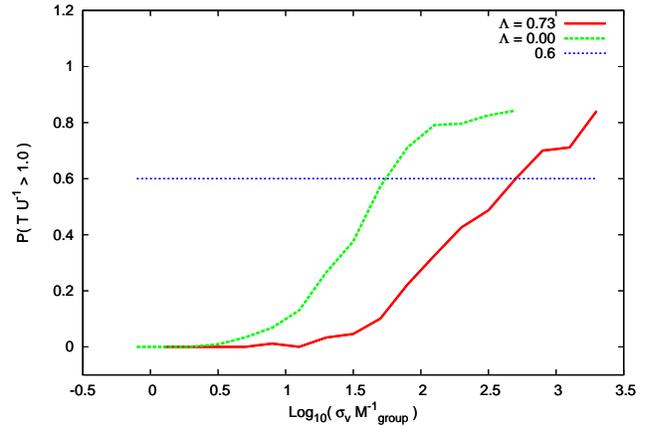}
\caption{Unbound probability ($P(T U^{-1} > 1.0)$) versus velocity dispersion 
($\sigma_{v}$) of a group when the $\Omega_{\Lambda} = 0.73$ model, the 
apparent magnitude limit of $20.0$, and a bin length of $0.2$ are adopted. 
$\sigma_{v}$M$^{-1}_{group}$ is in units of km s$^{-1}$ $(10^{12}h^{-1}~ M_{\odot})^{-1}$.}
	\label{fig15}
\end{figure}

\begin{figure}
\includegraphics[width=84mm]{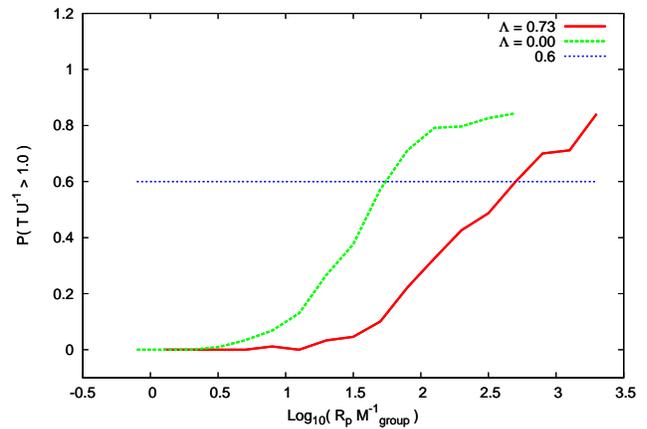}
\caption{Unbound probability ($P(T U^{-1} > 1.0)$) versus mean pairwise 
separation ($R_{p}$) when the $\Omega_{\Lambda} = 0.73$ model, the apparent 
magnitude limit of $20.0$, and a bin length of $0.25$ are adopted. 
$R_{p}$M$^{-1}_{group}$ is in units of Mpc $(10^{15}h^{-1}~M_{\odot})^{-1}$.}
	\label{fig16}
\end{figure}

The probability functions for the velocity dispersion $\sigma_{v}$ and the mean 
pairwise separation $R_{p}$, normalised to the group mass, are shown in Figures 
\ref{fig15} and \ref{fig16}. Figure \ref{fig15} shows that at values larger 
than $1.8$ (the horizontal dotted line in Figure \ref{fig15}) it is more 
probable that the groups are gravitationally unbound when the $\Omega_{\Lambda} 
= 0.73$ model is adopted. The same result can also be inferred from Figure 
\ref{fig9}, but with lower confidence.

The $\Omega_{\Lambda} = 0.0$ simulation gives a probability function which is 
comparable to the probability function of the $\Omega_{\Lambda} = 0.73$ model. 
It shows a similar linear growth as the probability function of the 
$\Omega_{\Lambda} = 0.73$ model. However, the function is shifted along the 
horizontal axis. This shifting originates from the variation of the group 
masses and velocity dispersions (see Figure \ref{fig13}). The change of the 
apparent magnitude limit does not have any significant effect on Figure 
\ref{fig15}. When the apparent magnitude limit of $13.2$ is adopted, a smaller 
number of haloes and groups are observed, which enlarges the variations between 
bins and gives a worse fit to a straight line.

The quantity $\sigma_{v}$M$^{-1}_{group}$, studied in Figure \ref{fig15}, has a 
loose connection to the kinetic energy $T$. This connection explains the fact 
that the lower normalized values of the quantity $\sigma_{v}$M$^{-1}_{group}$ 
gives gravitationally bound groups with a higher probability, and larger 
normalized values, loosely meaning the larger kinetic energies, gives unbound 
groups with a higher probability.

The probability function of the mean pairwise separation (Figure \ref{fig16}) 
shows a similar linear growth as the probability function of the velocity 
dispersion in Figure \ref{fig15}. The variation from bin to bin is somewhat 
larger in the probability function of the mean pairwise separation due to the 
fact that the mean pairwise separation is not strictly the size of the group 
but it includes projection effects. The quantity $R_{p}$M$^{-1}_{group}$, 
studied in Figure \ref{fig16}, is inversely proportional to the potential 
energy $U$ if the mean pairwise separation is identified as the real size of a 
group. The difference between Figures \ref{fig15} and \ref{fig16} is 
understandable, as the velocity dispersion and the mean pairwise separation are 
not strictly connected to each other, although some loose relation exists as 
the equation of the mean pairwise separation includes the mean group radial 
velocity.

The $\Omega_{\Lambda} = 0.90$ simulation show similar probability functions as 
the $\Omega_{\Lambda} = 0.0$ and the $\Omega_{\Lambda} = 0.73$ simulations. 
Figures \ref{fig15} and \ref{fig16} show that the $\Omega_{\Lambda}$ does not 
have any significant effect for the fraction of gravitationally  unbound 
groups. This result can also be inferred from Table \ref{tb:virialization}. The 
small effect is hardly surprising as the theoretical studies (see e.g. 
\citealt{LLPR}) have predicted that the $\Omega_{\Lambda}$ has little effect on 
the dynamics at the present epoch.

%Note, however, that the probability functions (Figures \ref{fig15} and \ref{fig16}) 
%and the method described above is difficult to apply in observations as the
%mass of the group must be obtain without the assumption of the virial theorem. 

\section[]{Conclusions}\label{summar}

%The main results of this wok are as follows:

We have shown that the $\Lambda$CDM cosmology can produce groups of dark matter 
haloes comparable to observations of groups of galaxies when the FOF algorithm 
based on that of \cite{HG82} is adopted and the dynamical properties of groups 
are studied. Our groups from cosmological simulations are, in general, in a 
moderate agreement with observations, although a straight K-S test fails in 
most cases. Our $\Omega_{\Lambda} = 0.73$ simulations are in satisfactory 
agreement with observations as the number densities of group properties are 
usually within 2$\sigma$ errors, or less, from the HG82 and the UZC-SSRS2 group 
abundances. The agreement between simulations and observations is good when the 
velocity dispersion and the 'observable' mass of groups is considered. In these 
cases the applied K-S test is approved when the high resolution 
$\Omega_{\Lambda} = 0.73$ simulation is considered. The moderate agreement 
between simulations and observational data suggests that gravitational force 
alone is sufficient in order to explain the dynamical properties of groups of 
galaxies.

We have also shown that in general about $20$ per cent of the groups of haloes 
generated with the algorithm presented in the HG82 are not gravitationally 
bound objects. The fraction of gravitationally bound groups of dark matter 
haloes varies with different values of the apparent magnitude limits. When the 
apparent magnitude limit is raised from the original $13.2$ to $20.0$, a larger 
number of spurious groups are found. The larger fraction of unbound 
groups with $m_{lim} = 20.0$ could be explained by the fact that more 
interlopers are included into groups, when the apparent magnitude limit is increased. 
However, this analysis is beyond the scope of this study.

In general, a larger number of 'rich' groups are found when the apparent 
magnitude limit is lowered. This originates from the fact that more light 
haloes at close proximity to more massive haloes become visible and those light 
haloes are included into the groups. When the magnitude limit is raised from 
the original value of $13.2$ to $12.0$, a slightly larger fraction of the 
groups are found to be gravitationally bound. In general, fewer groups (in 
absolutely number) are found and these groups are 'poorer'.

Small differences are found when the fractions of gravitationally bound 'poor' 
and 'rich' groups are studied. 'Rich' groups with more than four members are 
more often gravitationally bound than 'poorer' groups. This result agrees with 
previous ones (e.g. \citealt{RGPC}). Our results do not confirm the claim by 
\cite{RPG} who argued that 50 to 75 per cent of groups with three members are 
spurious. Our results show that $\sim 77$ per cent of groups containing 
only three members are gravitationally bound when the apparent magnitude limit 
of $13.2$ is adopted.

When the value of the cosmological constant $\Omega_{\Lambda}$ is varied, the 
fractions of unbound groups change only slightly. This is somewhat surprising as 
it would be intuitively expected that a larger value of the dark energy would 
lead to a greater number of groups that are not gravitationally bound. Some 
variation is observed when the fraction of gravitationally bound groups is 
studied as a function of the cosmological constant, but, in general, a significant 
number of groups remains unbound in all cases of $\Omega_{\Lambda}$.

When the values of the free parameters of the FOF algorithm are varied, the 
fraction of gravitationally bound groups can be raised from $\sim 81$ to $\sim 
83$ per cent. A greater difference is observed when the fraction of isolated 
haloes is studied. Varying the values of $D_{0}$ and $V_{0}$ makes a great 
difference, raising the fraction of isolated haloes from $\sim 42$ to $\sim 65$ 
per cent. In general, we do not find any significant difference in the fractions 
of gravitationally bound groups when different values of $D_{0}$ and $V_{0}$ or 
parameters of the Schechter luminosity function are adopted.
 
In observations, the crossing time of a group is often taken as an indicator 
of the virialization. We do not find any correlation between the virial ratio 
and the crossing time of a group. This result does not depend on the chosen 
value of the apparent magnitude limit of the search, or the cosmological model 
adopted. The lack of the correlation between these two variables calls into
question the crossing time as an estimator of the virialization.

\section*{Acknowledgements}

This work is part of the master's thesis of SMN at the University of Turku. SMN 
acknowledges the funding by the Finland's Academy of Sciences and Letters. SMN 
would like to thank Dr. Alexander Knebe for his cosmological N-body simulation 
code AMIGA, professor Gene Byrd for helpful suggestions, and the referee, 
Antonaldo Diaferio, for a number of invaluable corrections and suggestions.
The cosmological simulations were run at the CSC - Finnish IT center for
science.

%\appendix

\bsp

\label{lastpage}

\end{document}